# Kolmogorov Complexity, Causality And Spin

Dara O Shayda
April 2012

dara@lossofgenerality.com

## Abstract

A novel topological and computational method for 'motion' is described. Motion is constrained by inequalities in terms of Kolmogorov Complexity. Causality is obtained as the output of a high-pass filter, passing through only high values of Kolmogorov Complexity. Motion under the electromagnetic field described with immediate relationship with $G_2$ Holonomy group and its corresponding dense free 2-subgroup. Similar to Causality, Spin emerges as an immediate and inevitable consequence of high values of Kolmogorov Complexity. Consequently, the physical laws are nothing but a low-pass filter for small values of Kolmogorov Complexity.

## Motivation

In recent times, we are quite accustomed to viewing the universe via different filters e.g. Radio Telescopes or X-Ray Telescopes or Infrared CCD arrays and so on. We understand these varied outputs are 'the same' universe, however oddly different they might be rendered. If I showed you an optical photo of cosmos and told you: That is all there is, you would show me an X-Ray Chandra Satellite image and contradict. There is no ambiguity in any of these words.

What if there was a Complexity filter e.g. high-pass Kolmogorov Complexity filter which allowed only super high values to pass through. What would the output look like? Never ending Brownian motions. Basically a Foam Ocean. We see foam macroscopically and we see foam microscopically, hence self-similarity.

And what would be the Output of a low-pass Kolmogorov Complexity filter? Lines, points, circles and so on! Basically geometry. We see geometry macroscopically and we see geometry microscopically, hence self-similarity.

High-pass Kolmogorov Complexity filter only passes through irreversible motions, low-pass Kolmogorov Complexity passes though reversible motion.

Low-pass Kolmogorov Complexity renders a world of mirrors and mirror reflections, high-pass Kolmogorov Complexity renders a mirror-less reflection-less world.

However both are one and the same world!

# 1. Preliminaries

Let X be a topology and $W^*$ a Kleene star set of words composed from finitely many tokens (alphabets) W, define a function $Q$ for **quantization** of points in X:

$$Q: X \longrightarrow W^* \qquad (1.1)$$

**Remark 1.1**: *X can be any set without any topological structure, however its powerset always induces a natural topology in it.*

Borrowing from a technique in Real analysis $Q$ is extended to a continuous map $\overline{Q}$ to another topological space Y:

$$\overline{Q}: X \longrightarrow Y \supseteq W^* \qquad (1.2)$$

**Example 1.1**: Topological space 2-sphere or $S^2$ quantized in $\mathbb{R}^3$, quantized by assuming a small rational number $\theta$, with finite rational number approximations for Cos and Sin, concatenation of matrices corresponds to matrix multiplication. $GL(3, \mathbb{R})$ is a smooth manifold or topological space:

$$W = \left\{ \begin{pmatrix} 1 & 0 & 0 \\ 0 & \cos[\theta] & -\sin[\theta] \\ 0 & \sin[\theta] & \cos[\theta] \end{pmatrix}, \begin{pmatrix} \cos[\theta] & 0 & \sin[\theta] \\ 0 & 1 & 0 \\ -\sin[\theta] & 0 & \cos[\theta] \end{pmatrix}, \begin{pmatrix} \cos[\theta] & -\sin[\theta] & 0 \\ \sin[\theta] & \cos[\theta] & 0 \\ 0 & 0 & 1 \end{pmatrix} \right\} \quad , \quad GL(3, \mathbb{R}) \supseteq W^*$$

FIG 1.1

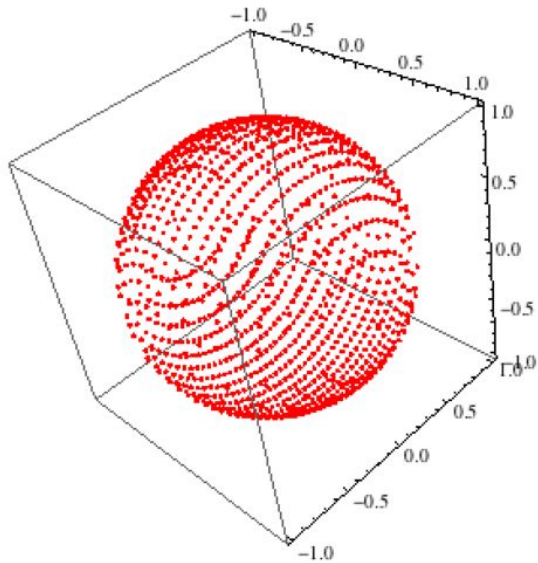

Assume there is a continuous path $c(t)$ connecting points A and B in X, and select finitely many points n on this path:

$$\{x_i\}_{i=1}^n \qquad c(0) = A, \ c(1) = B, \ x_i \in c$$

For each point on the said path find its corresponding quantized point in Y:

$$\{\overline{Q}(x_i)\}_{i=1}^n = \{w_i\}_{i=1}^n$$

$$\prod_{i=1}^n w_i = w_1 w_2 \ldots w_n = w$$

$\{\overline{Q}(x_i)\}_{i=1}^n = \{w_i\}_{i=1}^n \quad (1.3)$

Concatenate all the resulting words in Y:

$\prod_{i=1}^n w_i = w_1 w_2 \ldots w_n = w$

**Example 1.2**:

$\theta = 2\pi/100$

$$Rx = \begin{pmatrix} 1 & 0 & 0 \\ 0 & \cos[\frac{\pi}{50}] & -\sin[\frac{\pi}{50}] \\ 0 & \sin[\frac{\pi}{50}] & \cos[\frac{\pi}{50}] \end{pmatrix}$$

$$Ry = \begin{pmatrix} \cos[\frac{\pi}{50}] & 0 & \sin[\frac{\pi}{50}] \\ 0 & 1 & 0 \\ -\sin[\frac{\pi}{50}] & 0 & \cos[\frac{\pi}{50}] \end{pmatrix}$$

$$Rz = \begin{pmatrix} \cos[\frac{\pi}{50}] & -\sin[\frac{\pi}{50}] & 0 \\ \sin[\frac{\pi}{50}] & \cos[\frac{\pi}{50}] & 0 \\ 0 & 0 & 1 \end{pmatrix}$$

$W = \{Rx, Ry, Rz\}$

Pathword:

$\{w_i\}_{i=1}^n = \{(RxRyRz)^{2n}\}_{i=1}^{15}$

FIG 1.2

$$\{w_i\}_{i=1}^n = \{(RxRyRz)^{2n}\}_{i=1}^{15}$$

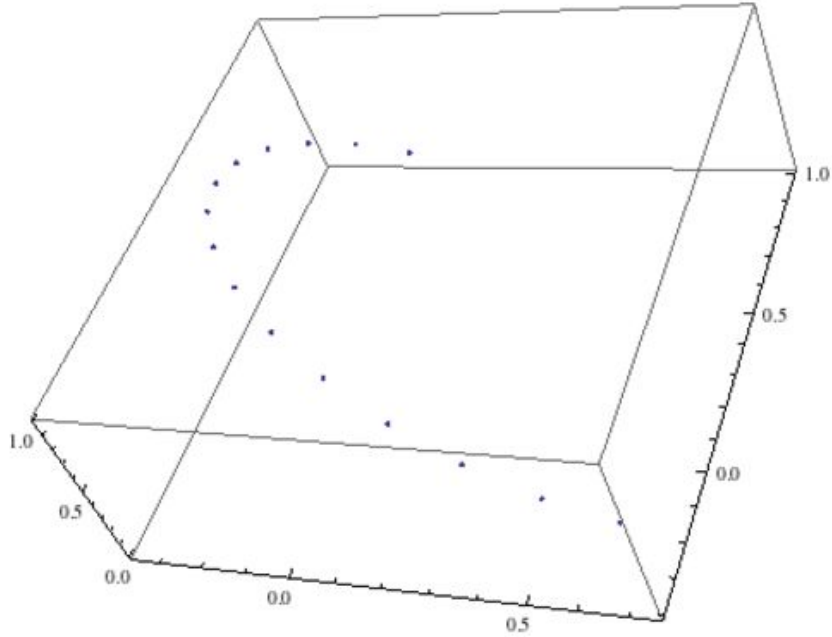

Loopword:

$$\{w_i\}_{i=1}^n = \{Rx^n\}_{i=1}^{100}$$

FIG 1.3

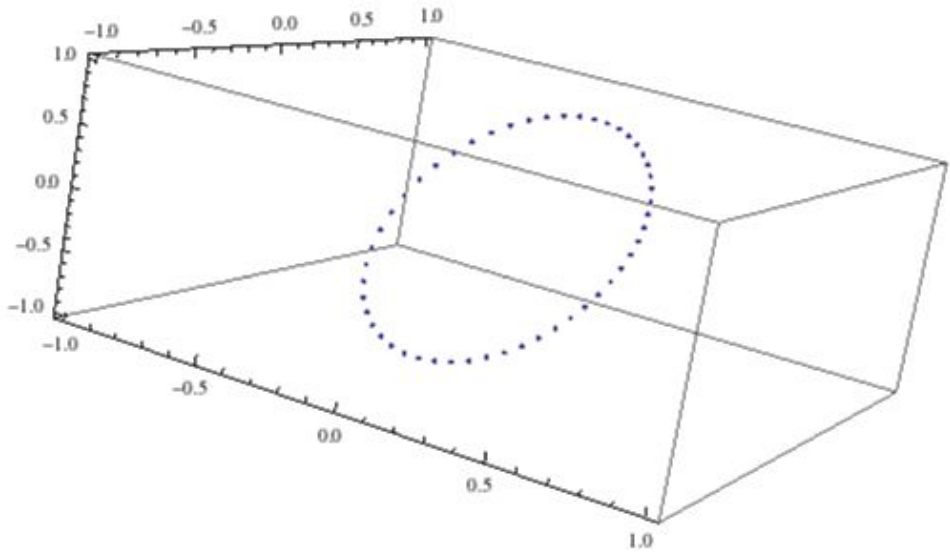

Let K(w) be the Kolmogorov Complexity (see Appendix A) of the concatenated word w. 'w' is called **Pathword** and in case of a loop called **Loopword**.

## 2. Idealization: Maxwellian Robot

Physical objects have been idealized as mathematical structures e.g. tuple of numbers or a set of something. Mathematical structures in and of themselves are quite limited. In order to progress with better forms of idealizations the author assumes objects in motions are idealized as robots! Thanks to the advent of Nano machines such idealization is natural and familiar.

By robot we mean a **mobile Turing machine**, which its inputs and outputs are coordinates of its motility in space-time.

Human mind describes the inner workings of nature by the objects surrounding him, in particular by the machines around him, as in the case of Maxwell imagined that space was filled with Ether that acted like 'cogs and gears' to move light from place to place:

This drawing was found in the 1861 paper of Maxwell [1] explaining the induction of the current by electromagnetic means, within a corpuscular medium. Three years later on in 1864 [2] he cognized that the matter media could be removed and reused the same ideas in 'vacua': "The we are obliged to admit that undulations are those of an aethereal substance and not of gross matter...".

FIG 2.1

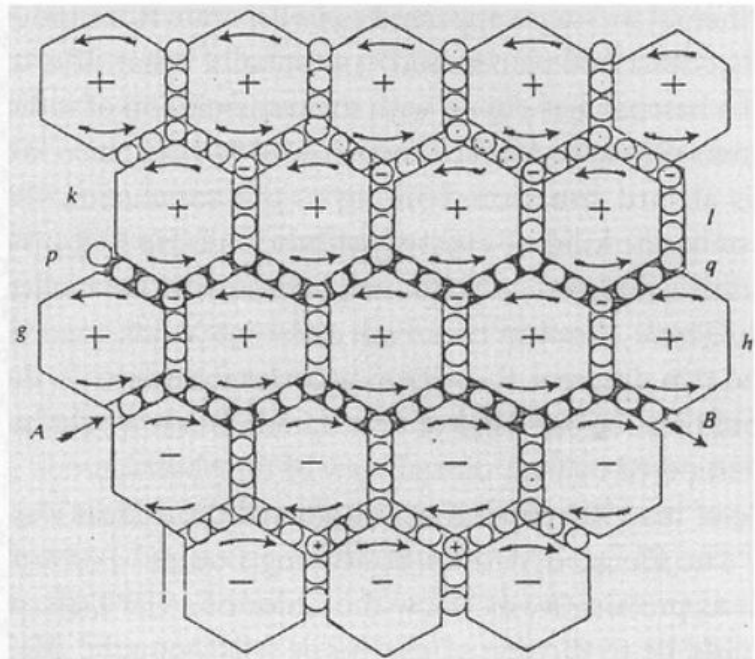

Today if anyone wrote a physics paper and used the 'cogs and wheels' as model for propagation of electromagnetic waves e.g. light, s/he would receive rejections from every publisher on the planet!

But it goes without saying that Maxwell's 'cogs and wheels', as a 'model', guided his thoughts and imagination to equations that perhaps are the best mathematical description of nature ever discovered by anyone.

We all know that space and light have no 'cogs and wheels', but the Maxwell's equations, derived from the 'cogs and wheels' model are still most valid computational and theoretical tools we have.

These arguments suffice for idealization of physical objects in particular particles as robots with limited memory and computation capacity. The author inspired by the cogs and wheels of Maxwell calls these mobile Turing machines **Maxwellian Robots**.

## 3. Motion: Self-Replication of Space

Maxwellian Robot is comprised of:

1. $\mathcal{U}$ a universal turning machine
2. p a self-delimiting program as input to $\mathcal{U}$
3. $c_0$ initial conditions e.g. position as additional input attached to p i.e. p, $c_0$ self-delimited
4. Program space in this Turing machine is limited i.e. the max size of a program is m

Motion in topological space X is modelled as an inverse function:

$$\left\{\overline{Q}^{-1}(w_i)\right\}_{i=1}^{n} \quad , \quad \overline{Q}: X \longrightarrow Y \supseteq W^* \qquad (3.1)$$

Where $\prod_{i=1}^{n} w_i = w_1 w_2 ... w_n = w$ is the output of the universal Turing machine ($w_i$ word concatenated with some separator scheme):

$$\mathcal{U}(p, c_0) = w_1 w_2 ... w_n = w \,, \quad w_i \in W^* \text{ and } |p| \leq m \quad (3.2)$$

w is the intended motion of the Maxwellian Robot, however there is deviation/fluctuation due to some external factor e.g. force or potential or other entities present:

$$E: W^* \longrightarrow W^* \qquad (3.3)$$

if $w_1 w_2 ... w_n = w$ then

$$E[w] = \prod_{i=1}^{n} E[w_i] \qquad (3.4)$$

This is an abuse of notation, so by $E[w]$ we always mean $\prod_{i=1}^{n} E[w_i]$ where $\mathcal{U}(p, c_0) = w_1 w_2 ... w_n$ . Moreover $E[w]$ is assumed to be self-delimiting (see Appendix A).

5. $\mathcal{U}$ is self-replicating i.e. it assembles or prints another copy of $\mathcal{U}$ , however with $p'$, $c'_0$ as possible different input (Mutation). No assumptions made on the synchronicity of the two copies.

**Remark 3.1**: *Self-replication is similar to system call fork() however there is no operating system to duplicate the executable code, in our case $\mathcal{U}$ itself replicates the copy. Best way of visualizing this form of self-replication is via self-printing or Quine programs [12]. The latter is a program that prints itself, with mutation e.g. in fork() the return value is the mutation, in our case the new $p'$, $c'_0$ is the mutation.*

If someone showed you a book and said: "it is just there, without any process of becoming, with all these meaningful words printed on each page", you would take up a stance against his idea.

'Space' is just there and has no process of generation! Similarly take a stance against this age old idea: Space needs to be generated, as in the case of a space of the book where each page is self-printed.

6. $\mathcal{U}$ is minimalistic i.e. it is smallest such universal Turing machine. This is so to reduce the size of the replication or

printing.

The output of the Maxwellian Robot is collection of points (subset) in some space. They self-replicate or self-print the space. Their motion is the generation of the space, or the printing of the space.

# 4. Causality: Direct Consequence of Large Kolmogorov Complexity

Imagine the robot traveling along the said points on a path as quantized by (3.1). Also assume the robot has limited onboard memory capacity m such that m is far less than the Kolmogorov Complexity of the finitely quantized path:

$$m \ll K(E[w]) \quad (4.1)$$

Q: Can the object/robot reverse the path?
A: No! It does not have enough memory to obtain a complete finite description of the path.

Therefore the path is irreversible for our object idealized as a robot. If we assume that X was time-space set of points and the points on the path were ordered in increasing order of time, this irreversibility is nothing but the Causality:

**Thesis 1**: *Collection of time-ordered paths with high Kolmogorov Complexity are Causal.*

Let's look at a robot with memory capacity close to the Kolmogorov Complexity:

$$m \approx K(E[w]) \quad (4.2)$$

Q: Can the object/robot reverse the path?
A: Yes! It does have enough memory to store a complete finite description of the path (compressed enough).

**Example 4.1**: Loopword in Example 1.2 if subjected to random fluctuations i.e. E simply added slight fluctuations to x and y and z coordinates, is a loop truly randomized then the reverse traversal is impossible.

FIG 4.1

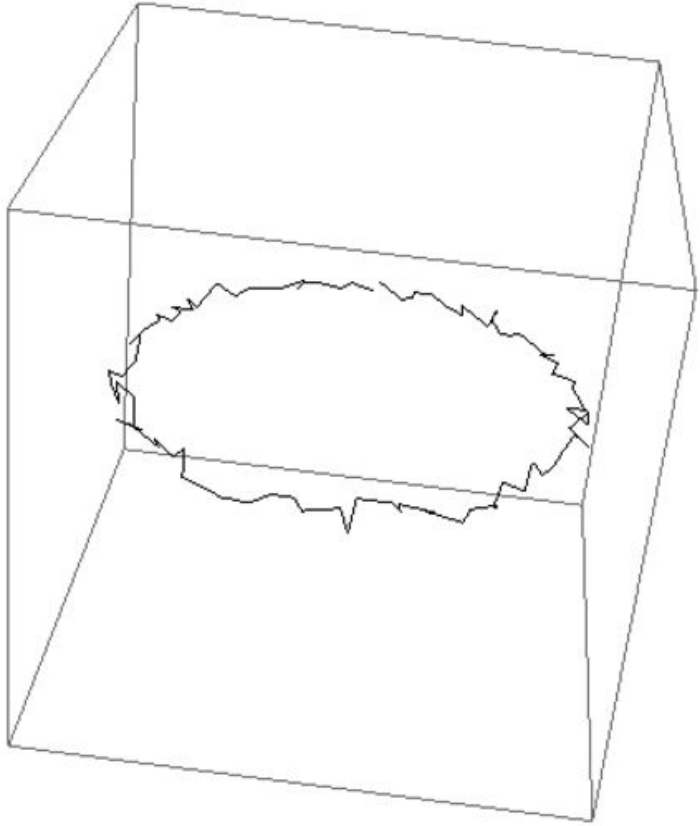

**Remark 4.1**: *Inequality (4.1) dictates that mostly matter can be observed moving along the positive direction of time. High Kolmogorov Complexity indicates the unavoidable Brownian motion of particles even in absence of all other matter or forces (i.e. vacuum) as was discovered by Feynman (see Discussion 3). Therefore:*

**Thesis 2**: *A Causal universe is brim-filled with Brownian motion or equivalently matter moving along the positive direction of time.*

# 5. Spin: Direct Consequence of Large Kolmogorov Complexity

Let's constrain a loop by the inequality (4.1):

$m \ll K(E[\text{Loopword}])$    (5.1)

Therefore the robot/object spinning around a loop in X cannot loop backwards since it does not have enough memory, and amazingly the concept of Spin emerges inevitably! As an immediate consequence of large Kolmogorov Complexity.

$$m \ll K(E[\text{Loopword}]) \qquad (5.1)$$

**Thesis 3**: *Collection of time-ordered loops with high Kolmogorov Complexity exhibit Spin.*

# 6. Electromagnetism: Y = $G_2$

Maxwell formulation of electricity and magnetism, and for most other such formulations, use the non-associative vector product or the cross product to calculate the force exerted on the charge [3]:

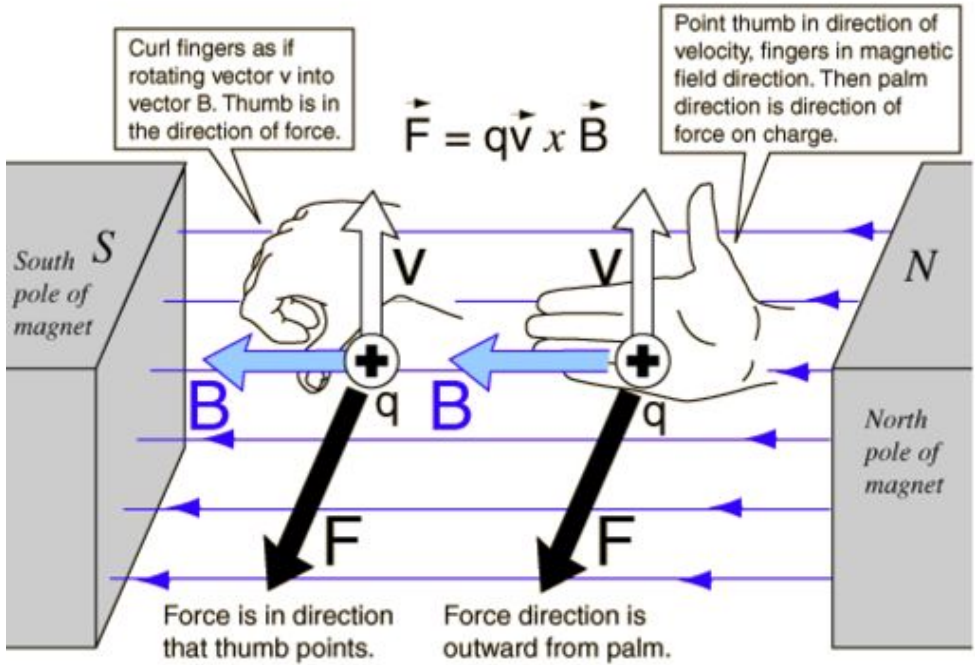

It was noted as a 'joke' by Feynman [4] in a failed attempt developing a new theory of electromagnetism, the algebra that supports non-associative cross product properties has maximum dimension of 7, later supported by the rigorous proofs given exclusively at Category level [5].

Therefore if we choose Riemannian manifold Y as the quantization space for points in normed/metric space X, then the dimension of Y or the dimension of its tangent space has to be maximum 7.

And if needed to quantize the loops in X, then the domain for quantization could be replaced by the Holonomy group of Y. Since Y has dimension 7 then there is only one Holonomy group for Y i.e. $G_2$ by Berger's classification theorem [6], consequently the quantization of the loops in X i.e. $c(S^1, X)$ or the continuous maps from circle to X, can be quantized in $G_2$:

$$\overline{Q}: c(S^1, X) \longrightarrow G_2 \supseteq W^* \qquad (6.1)$$

It is known that there is $W^*$ such that it is a dense free subgroup of $G_2$ generated by 2 generators $g_1$, $g_2$ and their inverses. (See Appendix B)

$$\phantom{G_2 \quad} W^*$$
$$G_2$$

$$\overline{Q}: c(S^1, X) \longrightarrow G_2 \supseteq W^* \qquad (6.1)$$

$W^*$ $G_2$ $g_1$ $g_2$

**Remark 3**: *$W^*$ is a non-smooth word concatenation free group with no other mathematical relations or structures, yet $G_2$ is a smooth 14 dimensional manifold! Indeed equation (6.1) is the bridge between the nowheredifferentiable world of Brownian motion that comprises our physical universe bridging to smooth mathematical structures which comprise our model of physical reality.*

The quantization of loops in X then can be formed by the finite strings/words:

$$\text{Loopword} = \prod_{i=1}^{n} g_j \quad j = 1, 2. \qquad (6.2)$$

Either a loop $l$ in X has a finite quantization (6.1) or has a sequence of such finite quantizations 'in limit' converging to the $Q(l)$ called the Loopword.

$G_2$ in particular preserves the cross products, therefore quantization based upon $W^* = <g_1, g_2, g_1^{-1}, g_2^{-1}>$ preserves the magnetic force (cross product) no matter what the aliasing due to the quantization. By aliasing it is meant the varying presentations for Loopword = $\prod_{i=1}^{n} g_j$ .

**Remark 4**: *The arguments clearly shows how and why $G_2$ has to be used for the quantization. In most physics papers they just pull the rabbit of $G_2$ out of an abstract hat and the reader cannot fathom why this particular holonomy. Indeed if the requirement of the non-associative cross product is removed, then any holonomy group would do.*

# 7. Physical Laws: Direct Consequence of Small Kolmogorov Complexity

In almost all literature the focus has been on the high Kolmogorov Complexity to study and understand the random and incompressible data. But what about the low Kolmogorov Complexity? What is the significance of reduced Kolmogorov Complexity:

**Manifesto**: *Physical Laws are small self-replicating programs generating finitely quantized paths or loops in space-time*

**Theorem 7.1**: *Any high Kolmogorov Complexity finitely quantized path or loop can be arbitrarily approximated by a much smaller Kolmogorov Complexity path or loop.*

**Proof**: Let's assume that there is a path or loop that with Kolmogorov Complexity that cannot be approximated arbitrarily by a lower Kolmogorov Complexity path or loop. Then there is a tube formed around the said path or loop, with its inside includes only very high Kolmogorov Complexity paths and loops. However a finite collection of points from inside this tube can be interpolated e.g. B-Splines of higher than 3rd degree and a parametric curve traverses through this tube with very low Kolmogorov Complexity, hence a contradiction.

FIG 7.1

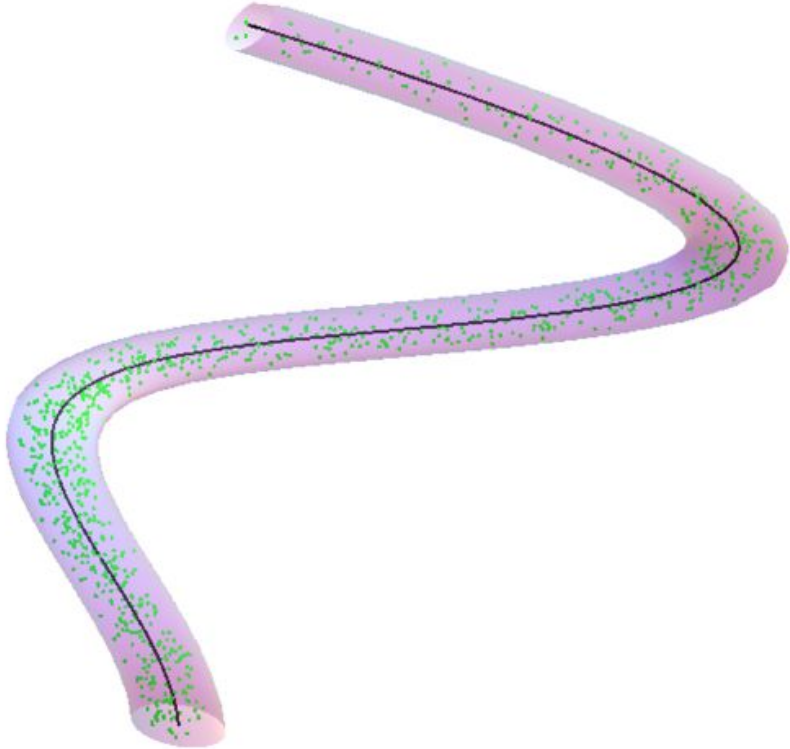

**Corollary 7.1**: *Low Kolmogorov Complexity physical laws can arbitrarily approximate high Kolmogorov Complexity paths and loops.*

**Remark 7.1**: *Since the Kolmogorov Complexity of physical law is small therefore the paths and loops can be reversed and hence the mirror symmetry of physical laws! However since the concept of 'small' is variant, then at some ranges of this smallness Kolmogorov Complexity can be big enough to cause symmetry breakage of the mirroring i.e. reverse path in time not possible.*

# 8. Self-Similarity: Direct Consequence of Reduced Kolmogorov Complexity

Direct consequence of this concept of Physical Law (i.e. low Kolmogorov Complexity) is similarities between microscopic and macroscopic (super large) physical structures abundant in nature. In other words, if the reduction of Kolmogorov Complexity is the emergence of the physical law then this reduction could occur no matter what magnitudes are the numbers in computation.

And the latter is true [11], as we can easily see the Spherical Harmonics i.e. the solutions to Shrodinger's equation have similar shapes to the stellar structures:

FIG 8.1

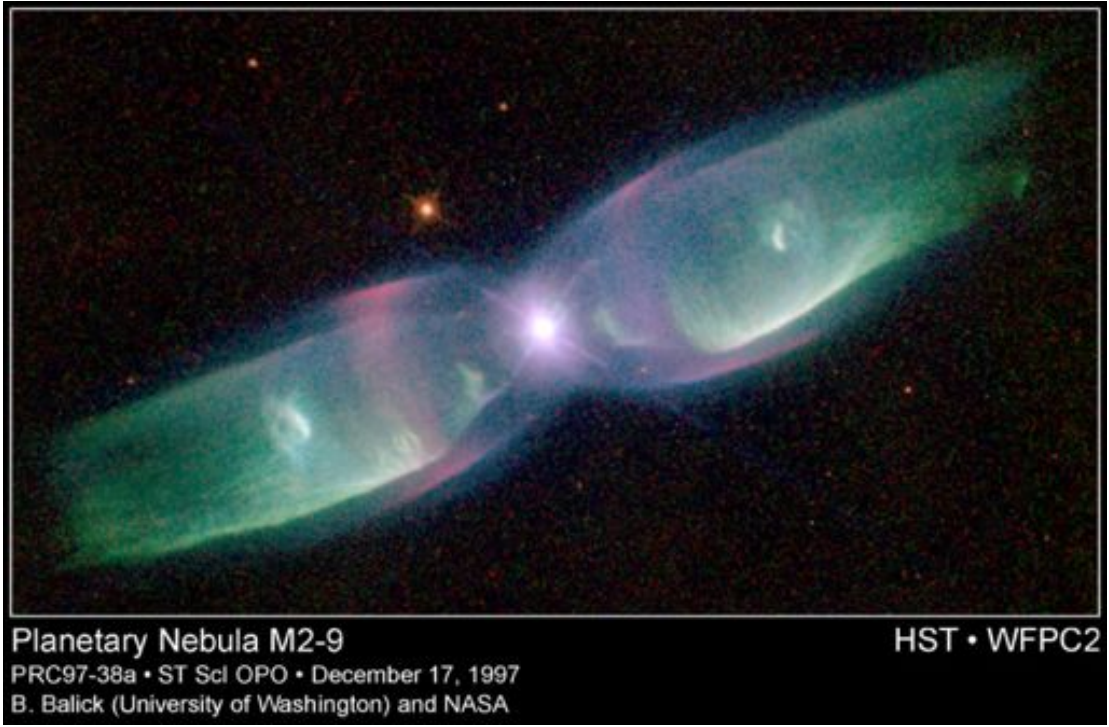

FIG 8.2

$$r = Y_5^2(\theta, \phi)$$

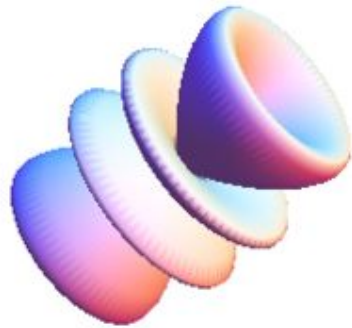

FIG 8.3

$$r = Y_7^1(\theta, \phi)$$

$$r = Y_7^1(\theta, \phi)$$

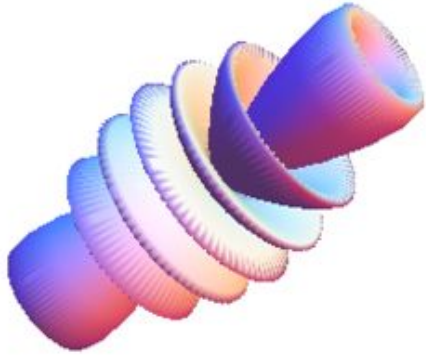

FIG 8.4

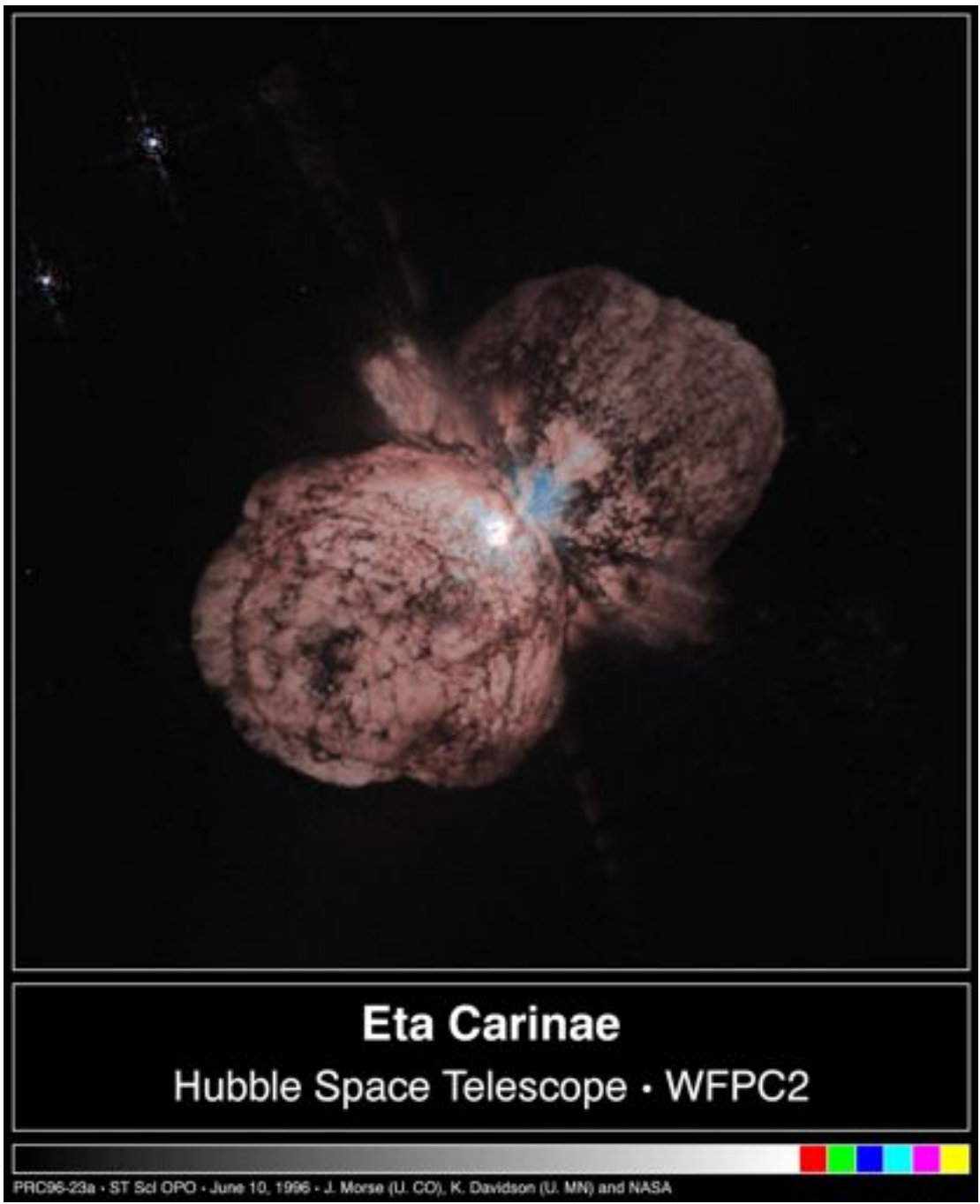

FIG 8.5

$$r = Y_4^2(\theta, \phi)$$

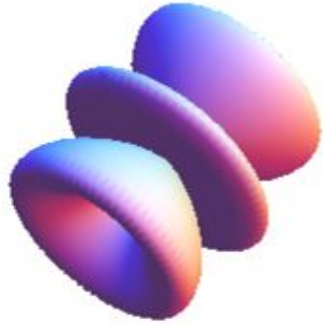

FIG 8.6

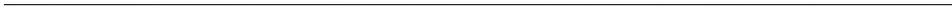

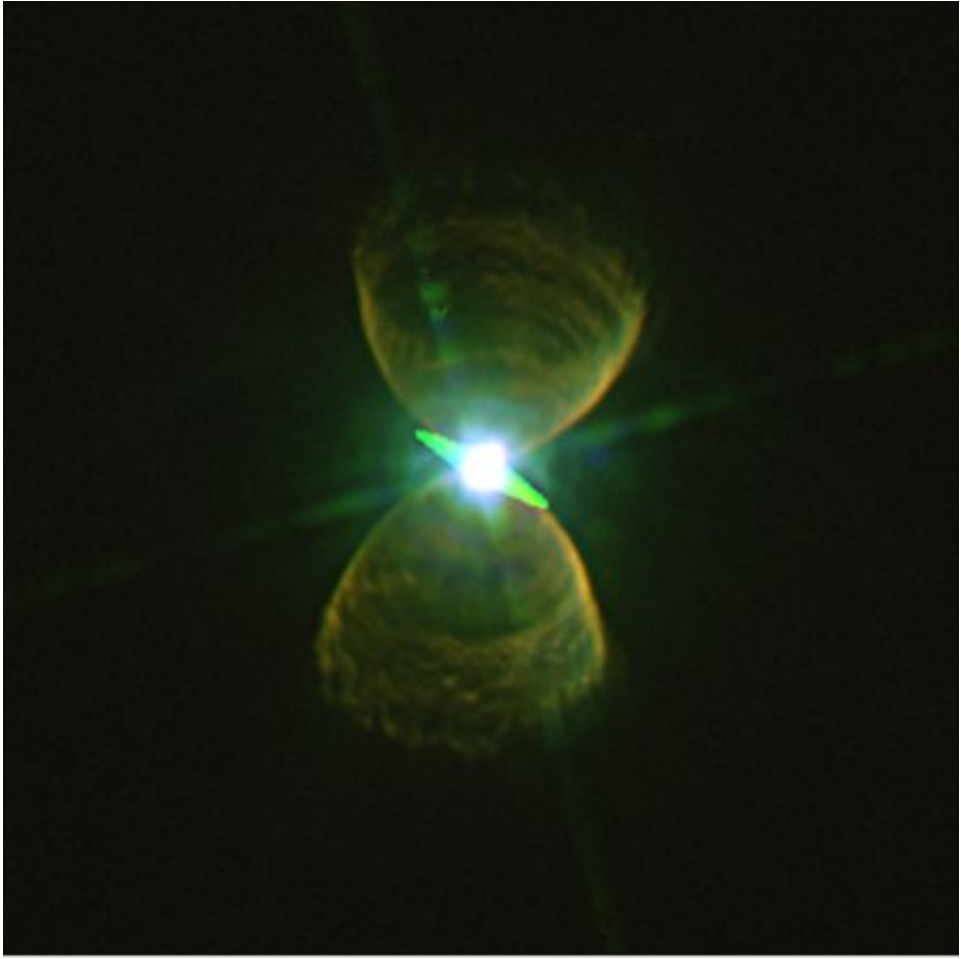

PN-G111.8-02.8 RGB F658N, F656N, F502N,PC1,FOV=17.6"

FIG 8.7

$$r = Y_5^4(\theta, \phi)$$

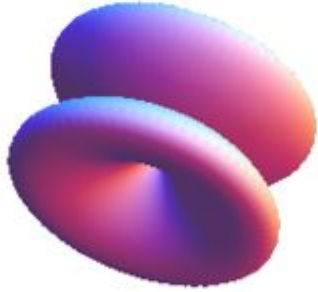

FIG 8.8

$$r = Y_3^0(\theta, \phi)$$

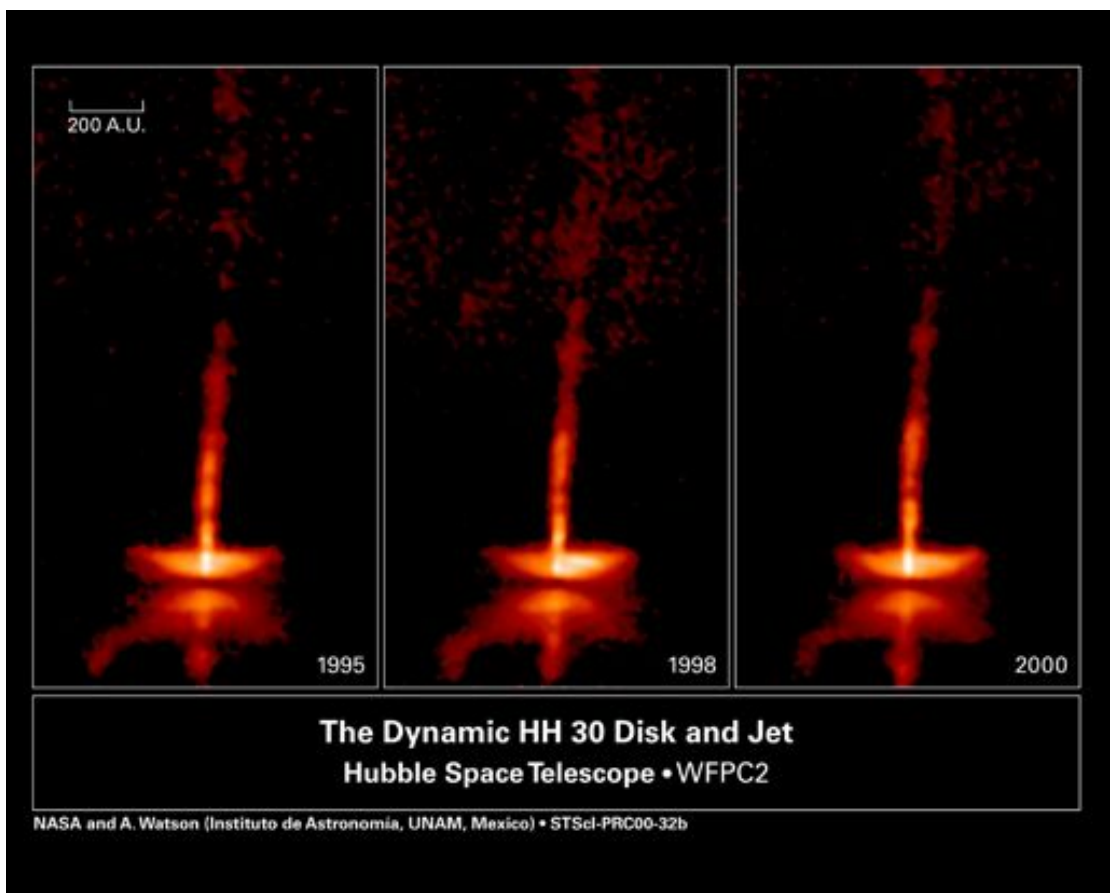

FIG 8.9

$$r = Y_3^0(\theta, \phi)$$

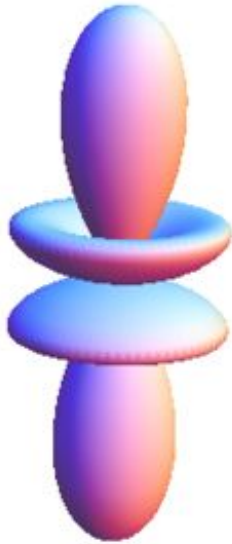

# Discussion

1. In this electromagnetic formulation the author did not use the Maxwell laws, for that matter any formulation of non-associative cross product algebras will abide by the same outcomes under similar assumptions about high/low Kolmogorov Complexity. The Kolmogorov Complexity clears the fog to see the real reasons why our observable world behaves in certain ways, while the pure mathematics might divert our attention to a wrong direction, and it does fail us. For example, high Kolmogorov Complexity paths in space-time are why we can only see the matter and little antimatter, since the backwards motion in time is not possible unless by drastic reduction of Kolmogorov Complexity.

Indeed the Kolmogorov Complexity acts as a high-pass filter.

2. Equation:

$$\overline{Q}: c(S^1, X) \longrightarrow G_2 \supseteq W^* \qquad (6.1)$$

Shows how the discreet $W^*$ is tied in with the smooth $G_2$. **The group theory is used to bridge between the nondifferentiable and differentiable**. Amazing fact is that the dense free subgroup of exceptional $G_2$ exists and therefore there is a natural choice of $W^*$.

3. Feynman' description of particle movement [9]:

"Typical paths of quantum-mechanical particle are highly irregular on a fine scale. Thus, though a mean velocity can be defined, no mean-square velocity exists at any point. In other words, the parts are nondifferentiable."

"If some average velocity is defined over a short time interval Δt, as, for example, [x(t + Δt) - x(t)]/Δt, the "mean" square value of this is -h/(imΔt). That is, the "mean" square value of the velocity averaged over a short time interval is finite, but is value becomes larger as the interval becomes shorter."

4. This formulation is free of ideas about particle anti-particle, it is all about reversing a word in a language.

5. Remark 4.1 clearly indicates that although the concept of Spin has been discovered for the charged particles in magnetic fields (Stern-Gerlach Experiment), Spin is a complexity concept, applicable to any kind of quantization.

6. Maxwellian Robots are a necessary cognitive gateway for thought processes to cognize the nature. Our theoretical physics is too mathematical and too little computational to gain new insights into the detailed workings of the nature.

# Appendix A

## Kolmogorov Complexity

We assume all strings and programs are binary coded.

**Definition A.1**: The Kolmogorov Complexity $C_\mathcal{U}(x)$ of a string x with respect to a universal computer (Turing Machine) $\mathcal{U}$ is defined as

$$C_\mathcal{U}(x) = \min_{p:\mathcal{U}(p) = x} l(p)$$

the minimum length program p in $\mathcal{U}$ which outputs x.

**Theorem A.1 (Universality of the Kolmogorov Complexity)**: *If $\mathcal{U}$ is a universal computer, then for any other computer $\mathcal{A}$ and all strings x,*

$$C_\mathcal{U}(x) \leq C_\mathcal{A}(x) + c_\mathcal{A}$$

where the constant $c_\mathcal{A}$ does not depend on x.

**Corollary A.1**: $\lim_{l(x) \to \infty} \frac{C_\mathcal{U}(x) - C_\mathcal{A}(x)}{l(x)} = 0$ *for any two universal computers.*

**Remark A.1**: *Therefore we drop the universal computer subscript and simply write C(x).*

**Definition A.2**: Self-delimiting string (or program) is a string or program including its own length.

**Definition A.3**: The Conditional or Prefix Kolmogorov Complexity of self-delimiting string x given string y is

$$K(x \mid y) = \min_{p:\mathcal{U}(p, y) = x} l(p)$$

The length of the shortest program that can compute both x and y and a way to tell them apart is

$$K(x, y) = \min_{p:\mathcal{U}(p) = xy} l(p)$$

$$K(x) \leq l(x) + 2 \log l(x), \quad K(x \mid l(x)) = l(x)$$

$$K(x, y) \leq K(x) + K(y)$$

$K(f(x)) \leq K(x) + K(f)$, f a computable function

$$K(x, y) = \min_{p:\mathcal{U}(p) = xy} l(p)$$

**Remark A.2**: *x, y can be thought of as concatenation of the strings with additional separation information.*

**Theorem A.2**: $K(x) \leq l(x) + 2 \log l(x), \quad K(x \mid l(x)) = l(x)$.

**Theorem A.3**: $K(x, y) \leq K(x) + K(y)$.

**Theorem A.4**: $K(f(x)) \leq K(x) + K(f)$ , $f$ a computble function

**Definition A.3**: Two strings x and y are independent if $K(x \mid y) = K(x) + K(y)$.

# Appendix B

## $G_2$ Holonomy Group

Given a connection $\nabla$ on a Riemannian manifold (M, g), any path maps a vector in the tangent space of one point to another vector in the tangent space of the endpoint i.e. a parallel transport of the local tangent vector to the endpoint tangent vector. Therefore each parallel transport along a path induces a linear map on the tangent space. The connection is arbitrary, basically it smoothly maps the differentials at a point to the nearby points' differentials.

The source tangent vector (green arrow) below is parallel transport to the target tangent vector (red arrow), which induces a linear transformation from the source tangent space to the target tangent space.

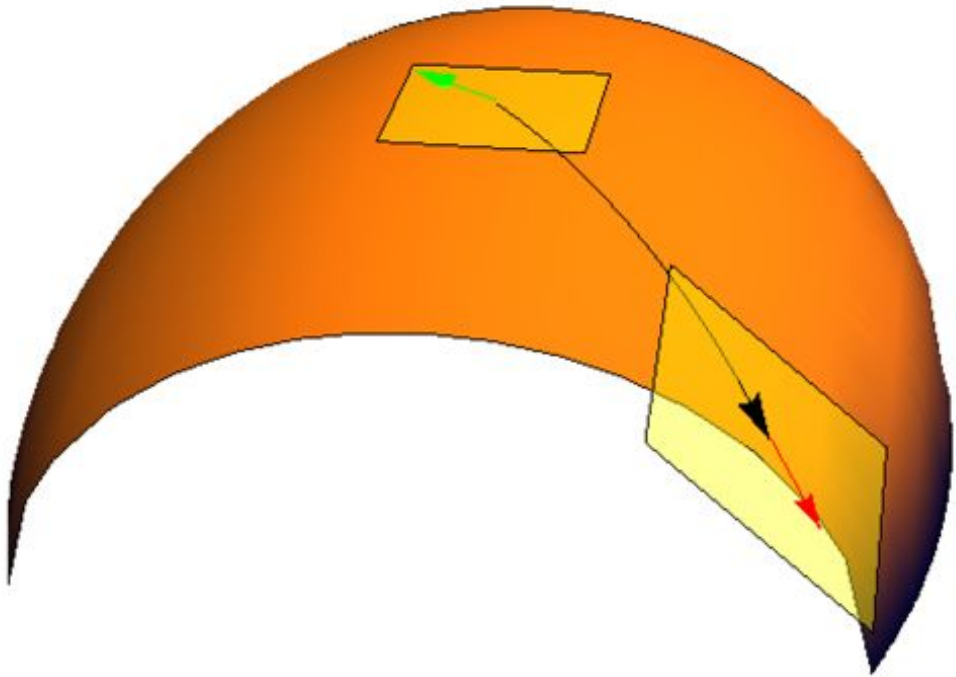

If the path is a loop, then the resulting linear maps form a group, indeed a subgroup of GL (T) where T is the tangent space of (M, g). This group is called the Holonomy group of (M, g) or Hol ($\nabla$), up to conjugacy.

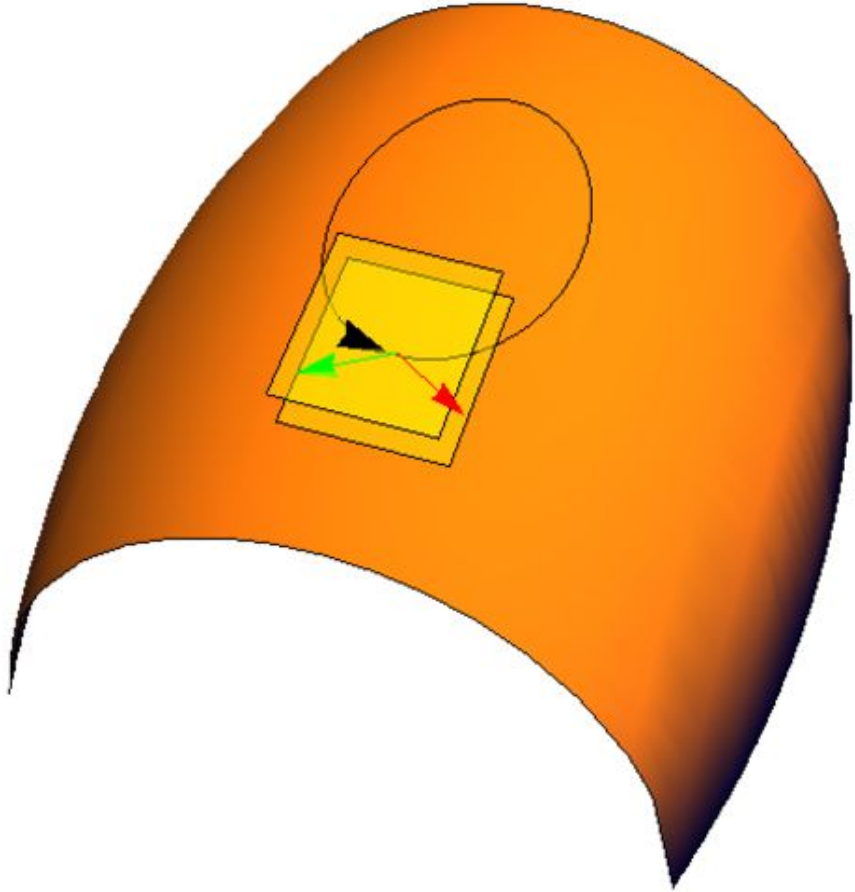

For dimension 7 Riemannian manifolds the Holonomy group is the exceptional (i.e. exclusive to dimension 7) $G_2$.

Properties of $G_2$ [10]:

1. $G_2$ itself is a (matrix) manifold of dimension 14.
2. $G_2$ is the group of automorphisms of Octonions to Octonions:

$G_2 = \text{Aut}(\mathbb{O}) = \{g \in \text{GL}(\mathbb{O}) \cong \text{GL}_8(\mathbb{R}) \mid g(xy) = g(x)\,g(y)\}$

3. $G_2$ preserves the cross product of Octonions [10]:

$x \times y = [x, y]/2$ then $g(x \times y) = g(x) \times g(y) \quad \forall\, g \in G_2$

Matrix representation of $G_2$ can either be a 8×8 or 7×7 matrix depending on including or excluding the Real part of the Octonions.

Samples of 7×7 matrix representations in $G_2$:

$$\begin{pmatrix} 1 & 0 & 0 & 0 & 0 & 0 & 0 \\ 0 & \cos[x1] & \sin[x1] & 0 & 0 & 0 & 0 \\ 0 & -\sin[x1] & \cos[x1] & 0 & 0 & 0 & 0 \\ 0 & 0 & 0 & \cos[x1] & -\sin[x1] & 0 & 0 \\ 0 & 0 & 0 & \sin[x1] & \cos[x1] & 0 & 0 \\ 0 & 0 & 0 & 0 & 0 & 1 & 0 \\ 0 & 0 & 0 & 0 & 0 & 0 & 1 \end{pmatrix}$$

$G_2$

$$g = \begin{pmatrix} 1 & 0 & 0 & 0 & 0 & 0 & 0 \\ 0 & \cos[x1] & \sin[x1] & 0 & 0 & 0 & 0 \\ 0 & -\sin[x1] & \cos[x1] & 0 & 0 & 0 & 0 \\ 0 & 0 & 0 & \cos[x1] & -\sin[x1] & 0 & 0 \\ 0 & 0 & 0 & \sin[x1] & \cos[x1] & 0 & 0 \\ 0 & 0 & 0 & 0 & 0 & 1 & 0 \\ 0 & 0 & 0 & 0 & 0 & 0 & 1 \end{pmatrix}$$

$g(i\,j) = g(i)\,g(j)$

or

$$g = \begin{pmatrix} 1 & 0 & 0 & 0 & 0 & 0 & 0 \\ 0 & 1 & 0 & 0 & 0 & 0 & 0 \\ 0 & 0 & 1 & 0 & 0 & 0 & 0 \\ 0 & 0 & 0 & \cos[y1] & 0 & 0 & \sin[y1] \\ 0 & 0 & 0 & 0 & \cos[y1] & \sin[y1] & 0 \\ 0 & 0 & 0 & 0 & -\sin[y1] & \cos[y1] & 0 \\ 0 & 0 & 0 & -\sin[y1] & 0 & 0 & \cos[y1] \end{pmatrix}$$

$g(e_4 \times e_7) = g(e_4) \times g(e_7)$